\pdfoutput=1

\documentclass[aps,prl,reprint,amsmath,amsfonts,showkeys]{revtex4-2}
\usepackage[colorlinks,allcolors=blue]{hyperref}
\usepackage[T1]{fontenc}
\usepackage[USenglish]{babel}
\usepackage{graphicx}
\usepackage{bm}

\begin{document}
\title{Exotic Hidden-heavy Hadrons and Where to Find Them}
\author{Eric Braaten}
\email{braaten.1@osu.edu}
\affiliation{Department of Physics,
	         The Ohio State University, Columbus, OH\ 43210, USA}
\author{Roberto Bruschini}
\email{bruschini.1@osu.edu}
\affiliation{Department of Physics,
         The Ohio State University, Columbus, OH\ 43210, USA}

\begin{abstract}
The Born-Oppenheimer potentials for QCD with light quarks include adjoint-hadron potentials that are repulsive at short distances and heavy-hadron-pair potentials that approach thresholds at  large distances. The adjoint-hadron potentials must connect smoothly to the heavy-hadron-pair potentials at intermediate distances. We identify exotic hidden-heavy hadrons as bound states and resonances in adjoint-hadron potentials that cross below a heavy-hadron-pair threshold before approaching it. This explains why many exotic hidden-charm and hidden-bottom hadrons have energies near heavy-hadron-pair thresholds. The remarkable properties of some  exotic hidden-heavy mesons can be explained by  fine tunings of  adjoint-meson energies in QCD.
\end{abstract}


\maketitle

\textbf{Introduction.}
Dozens of  exotic heavy hadrons have been discovered since the beginning of the 21\textsuperscript{st} century
\cite{Bra20}.
Most of them are hidden-heavy hadrons whose
constituents  include a charm or bottom quark-antiquark ($c \bar{c}$ or $b \bar{b}$) pair.
By a recent count, they  include as many as
43 $c \bar{c}$ tetraquark mesons, 5 $b \bar{b}$ tetraquark mesons, and 5 $c \bar{c}$ pentaquark baryons \cite{Leb23b}.
Their discoveries challenge our understanding of quantum chromodynamics (QCD) 
and other strongly interacting gauge theories.

Most models for exotic heavy hadrons have tenuous connections to the fundamental theory QCD.
They have multiple parameters with unknown relations to the QCD coupling constant $\alpha_\text{s}$ and the quark masses. 
Molecular models  whose constituents are color-singlet hadrons
are motivated by many exotic heavy hadrons having energies near heavy-hadron-pair thresholds. 
Models whose constituents have color charges predict an explosion in the number of exotic heavy hadrons, 
making it easy to accommodate new discoveries by overlooking the many 
predictions of unobserved states.

\begin{figure}[t]
\includegraphics[width=\columnwidth]{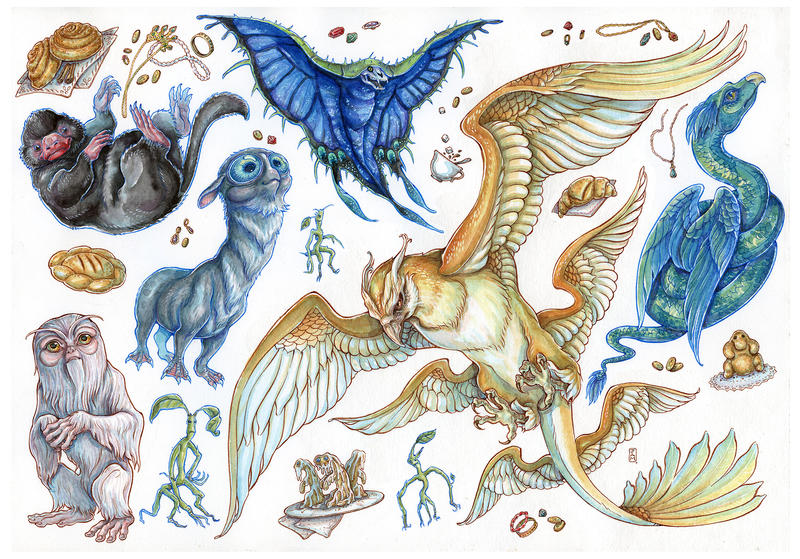}
\caption{Exotic hidden-heavy hadrons and other fantastic beasts (image by Jam-Di, \href{https://www.deviantart.com/jam-di}{deviantart.com/jam-di}).}
\label{fig:beasts}
\end{figure}

The Born-Oppenheimer (B\nobreakdash-O) approach to multi-heavy hadrons is based firmly on QCD.
It has been developed into an effective field theory for multi-heavy hadrons called BOEFT  \cite{Ber15,On17,Bra18a,Sot20a}.
The B\nobreakdash-O approach separates the problem into two steps:
1) calculating B\nobreakdash-O potentials using lattice QCD,
2) solving the Schr\"odinger equation for heavy quarks and antiquarks in the B\nobreakdash-O potentials.
The B\nobreakdash-O approach provides a unified framework for describing all multi-heavy hadrons
\cite{Bra18a,Sot20a,Braa24,Ber24,Bru24}.
Although theoretically compelling, it has not yet explained the pattern of the observed exotic heavy hadrons
or made compelling predictions of new ones before their discoveries.

In this paper, we emphasize that the B\nobreakdash-O potentials for hidden-heavy hadrons
that approach heavy-hadron-pair thresholds at large distances must, with the exception of the ground-state  potential, 
connect smoothly to repulsive $1/r$ potentials at short distances.
This severely reduces the B\nobreakdash-O  channels that can support bound states and resonances.
It explains why many exotic hidden-heavy hadrons have energies near heavy-hadron-pair thresholds.
We also identify the fine tunings of QCD responsible for the remarkable properties of some exotic hidden-heavy mesons.

\textbf{Static Color Sources.}
The  B\nobreakdash-O approximation exploits the large mass of a heavy quark 
compared to the momentum scales of gluons and light quarks.
The B\nobreakdash-O potentials for hidden-heavy hadrons are discrete energies 
in the spectrum of QCD with static $\bm{3}$ and $\bm{3}^\ast$ color sources separated by a distance $r$.
The sources reduce the rotational symmetry to a cylindrical subgroup
with quantum number $\lambda$  that is integer or half-integer.
They reduce the  parity  and charge-conjugation symmetries with quantum numbers $P$ and $C$ to their product $CP$.
There is also  a reflection symmetry through a plane containing the two sources with quantum number $R$.
The B\nobreakdash-O  symmetries are traditionally denoted $\Lambda_\eta^\epsilon$, 
where  $\Lambda = \lvert\lambda\rvert$ (or $\Sigma,\Pi,\Delta, \dotsc$ if $\lvert\lambda\rvert$ is $0,1,2, \dotsc$),
$\eta=g,u$ if $CP = +1,-1$, and $\epsilon=+,-$ if $R=+1,-1$.
If $\Lambda$ is not $\Sigma$, $\epsilon$ is omitted.
The B\nobreakdash-O potentials are also labeled using approximate light-quark flavor symmetries, such as isospin.

Calculating B\nobreakdash-O potentials at large $r$ and small $r$ reduces to the simpler problem 
of  QCD  with a single static color source.
Its spectrum is determined up to an additive constant that depends on the color source.
The discrete states are labeled by quantum numbers $J^{PC}$, where $J$ is the angular momentum.
They  are also labeled using light-quark flavor symmetries.
A discrete state  bound to a static $\bm{3}$ or $\bm{3}^\ast$ color source is called a \emph{$\bm{3}$-hadron} or \emph{$\bm{3}^\ast$-hadron}.
A discrete state bound to a static $\bm{8}$ color source is called an \emph{adjoint hadron} or an \emph{$\bm{8}$-hadron}.

As $r \to \infty$, confinement and the cluster decomposition requires the finite-energy states of QCD with $\bm{3}$ and  $\bm{3}^\ast$ sources to be pairs of a $\bm{3}$-hadron and a $\bm{3}^\ast$-hadron.
We refer to the corresponding B\nobreakdash-O potentials at large $r$ as  ($\bm{3}$+$\bm{3}^\ast$)-potentials.
As $r \to 0$, the $\bm{3}$ and $\bm{3}^\ast$ sources reduce to
a linear combination of a color-singlet ($\bm{1}$) source and an $\bm{8}$ source.
QCD with a $\bm{1}$ source, which is the same as no source, has only one discrete state:
the QCD vacuum with B\nobreakdash-O quantum numbers $\Sigma_g^+$.
The corresponding $\bm{1}$-potential is an attractive $1/r$ potential at small $r$.
We refer to a B\nobreakdash-O potential at small $r$ 
whose QCD state approximates an $\bm{8}$-hadron as an $\bm{8}$-potential.
As $r \to 0$, an $\bm{8}$-potential approaches a repulsive $1/r$ potential shifted by the $\bm{8}$-hadron energy. 

The spectrum of QCD with two static color sources must be a smooth function of $r$.
This provides surprisingly strong constraints on the B\nobreakdash-O potentials for hidden-heavy hadrons.
 The ground-state $\Sigma_g^+$ potential  connects a ($\bm{3}$+$\bm{3}^\ast$)-potential 
at large $r$ to the $\bm{1}$-potential at small $r$.
All other ($\bm{3}$+$\bm{3}^\ast$)-potentials at large $r$ must connect to $\bm{8}$-potentials at small $r$.
For double-heavy hadrons with two heavy quarks, the $(\bm{3} \!+\! \bm{3})$-potentials at large $r$ in lattice QCD
connect at small $r$ to either attractive $\bm{3}^\ast$-potentials 
or repulsive $\bm{6}$-potentials \cite{Bic16}.
The analogous fact for hidden-heavy hadrons has only been recognized recently \cite{Ber24}.

There have been few calculations of the spectrum of  $\bm{8}$-hadrons using lattice QCD.
The first calculations for gluelumps  ($\bm{8}$-mesons that are  $\mathrm{SU(3)}$-flavor singlets)
by Foster and Michael in 1998 used pure $\mathrm{SU(3)}$ gauge theory \cite{Fos99}.
The ground-state gluelump is $J^{PC}= 1^{+-}$. 
The spectrum has recently been calculated more accurately \cite{Her24}.
There are also lattice calculations with 2+1 flavors of light quarks \cite{Mar14}. 
The  spectrum of $\bm{8}$-mesons besides gluelumps has only been calculated using $\mathrm{SU(3)}$ gauge theory 
with light valence quarks \cite{Fos99}.
The lowest-energy $\bm{8}$-mesons are $1^{--}$ and $0^{-+}$. 
Various models for QCD give more detailed predictions for the spectrum of $\bm{8}$-hadrons,
as summarized in Ref.~\cite{ATLAS19review}.

\textbf{History of B-O Potentials.}
The first quantitative QCD result from lattice gauge theory was the $\Sigma_g^+$ potential
in pure $\mathrm{SU(3)}$ gauge theory in 1984 \cite{Sta84}.
This confining potential is an attractive $1/r$ potential at small $r$ and increases linearly at large $r$.
The ground-state $\Sigma_g^+$ potential in QCD with light quarks
must cross over at large $r$ to a constant  equal to twice the energy of the ground-state $\bm{3}$-meson.
Bali \textit{et al.}\ showed in 1995 that this occurs through a narrow avoided crossing with an excited $\Sigma_g^{+\prime}$ potential  
that crosses over from approximately constant to increasing linearly with $r$ \cite{Bal05}.
This suggested that QCD with light quarks has  additional ($\bm{3}$+$\bm{3}^\ast$)-potentials at large $r$ 
for every pair of a $\bm{3}$-hadron and a $\bm{3}^\ast$-hadron.

The B\nobreakdash-O approximation for QCD was pioneered in 1999 by Juge, Kuti, and Morningstar,
who applied it to quarkonium hybrid mesons  \cite{Jug99}.
They showed that pure $\mathrm{SU(3)}$ gauge theory also has higher potentials with other B\nobreakdash-O quantum numbers \cite{Jug03}. 
They are confining potentials that increase linearly at large $r$.
The development of pNRQCD revealed that they are $\bm{8}$-potentials
at small $r$ that differ by  gluelump energy differences  \cite{Bra00}.

The discoveries of $T_{b\bar{b}1}^+(10610)$ and $T_{b\bar{b}1}^+(10650)$ 
($Z_b$ and $Z_b^\prime$) in 2011 \cite{BELLE12}
and $T_{c\bar{c}1}^+(3900)$ in 2013 \cite{BESIII13,BELLE13}
 motivated the realization that QCD has additional $\bm{8}$-potentials
for every $\bm{8}$-hadron whose flavor is not $\mathrm{SU(3)}$-singlet \cite{Braa13,Braa14}.
At small $r$, they approach a repulsive $1/r$ potential shifted by the energy of the $\bm{8}$-hadron.
Charged quarkonium states were identified as bound states in these potentials,
which were assumed to increase linearly at large $r$.
New confining potentials for every  $\bm{8}$-hadron imply
an explosion in the number of exotic hidden-heavy hadrons.

\begin{figure}
\centering
\includegraphics{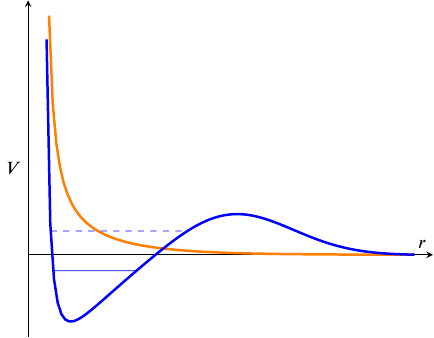}
\caption{%
An $\bm{8}$-potential at short distances
can approach a ($\bm{3}$+$\bm{3}^\ast$)-potential at long distances
by decreasng monothonically (upper curve at small $r$). It may also decrease below the
($\bm{3}$+$\bm{3}^\ast$)-threshold before approaching it from below or from above (lower curve at small $r$), in which case it may support a bound state (solid horizontal line) or a resonance (dashed horizontal line).}
\label{fig:potentials}
\end{figure}

The continuity constraint that an $\bm{8}$-potential at small $r$ 
connects to a ($\bm{3}$+$\bm{3}^\ast$)-potential at large $r$ 
severely limits the B\nobreakdash-O potentials capable of supporting bound states.
An $\bm{8}$-potential at small $r$ decreases with increasing $r$.
The simplest possibility is that it decreases monotonically to the
 ($\bm{3}$+$\bm{3}^\ast$)-threshold,
in which case it cannot support bound states.
The second-simplest possibility is that it crosses below the  ($\bm{3}$+$\bm{3}^\ast$)-threshold 
and then approaches it from below, in which case it can support bound states.
The third-simplest possibility is that it crosses below the ($\bm{3}$+$\bm{3}^\ast$)-threshold,
crosses it again, and then approaches it from above, in which case it can also support resonances.
The simplest and third-simplest possibilities are illustrated in Fig.~\ref{fig:potentials}.
We identify these bound states and resonances as exotic hidden-heavy hadrons.
This explains why the masses of most observed exotic hidden-heavy hadrons are near heavy-hadron-pair thresholds.
Since energy splittings between $\bm{8}$-hadrons are typically larger than those between pairs of a
$\bm{3}$-hadron and a $\bm{3}^\ast$-hadron,
most $\bm{8}$-potentials approach a $(\bm{3} \!+\! \bm{3}^\ast)$-threshold from above.  
Only those associated with the lowest-energy $\bm{8}$-hadrons 
cross below the ($\bm{3}$+$\bm{3}^\ast$)-threshold before approaching it.
This avoids an explosion in the number of exotic hidden-heavy hadrons.
The thresholds with exotic hidden-heavy hadrons nearby can be  predicted most easily 
by using lattice QCD to calculate the spectrum of $\bm{8}$-hadrons.

\textbf{Model B-O potentials.}
Lattice calculations in $\mathrm{SU(3)}$ gauge theory with valence quarks suggest that 
the lowest-energy $\bm{8}$-mesons are $J^{PC}=1^{--}$  and $0^{-+}$ \cite{Fos99}.
The B\nobreakdash-O  potentials associated with a $1^{--}$ $\bm{8}$-meson are $\Sigma_g^{+\prime}$ and $\Pi_g$.
The B\nobreakdash-O potential associated with a $0^{-+}$ $\bm{8}$-meson   is $\Sigma_u^-$.
The lowest-energy multiplets in these B\nobreakdash-O  potentials are given in Table~I of Ref.~\cite{Ber24}.
For the $1^{--}$ $\bm{8}$-meson, the ground-state multiplet is a quartet consisting of a heavy-quark-pair spin-singlet ($S=0$)  $1^{+-}$ 
and a spin-triplet ($S=1$) $(0,1,2)^{++}$.
For the $0^{-+}$ $\bm{8}$-meson,  
the ground-state multiplet is a doublet consisting of a spin-singlet  $0^{++}$ and a spin-triplet  $1^{+-}$. 
Note that $0^{++}$  and $1^{+-}$ also appear in the ground-state multiplet for the $1^{--}$ $\bm{8}$-meson.

For the purposes of illustration, we use simple models for the $\Sigma_g^{+\prime}$, $\Pi_g$, and  $\Sigma_u^-$ potentials. 
We take the zero of energy to be the spin-weighted isospin-averaged heavy-meson-pair threshold:
$E_{D \bar{D}} =3946$\,MeV for charm mesons and $E_{B \bar{B}} =10627$\,MeV for bottom mesons.
Our model for the $\Lambda_\eta^\epsilon$ potential associated with the $J^{PC}$ $\bm{8}$-meson with isospin $I$ is
\begin{equation}
V_{\Lambda_\eta^\epsilon}^{(I)}(r) =
\Biggl\{
\!
\begin{array}{ll}
\kappa_8 / r +  E_{J^{PC}}^{(I)} + A_{\Lambda_\eta^\epsilon}^{(I)}\, r^2  &
\text{if $r < R_{\Lambda_\eta^\epsilon}^{(I)}$,} \\
B_{\Lambda_\eta^\epsilon}^{(I)}\,e^{-r/d}  &
\text{if $r > R_{\Lambda_\eta^\epsilon}^{(I)}$,}
\end{array}
\label{V-Lambdaetaepsilon}
\end{equation}
where $R_{\Lambda_\eta^\epsilon}^{(I)}$ and $B_{\Lambda_\eta^\epsilon}^{(I)}$ 
are determined by continuity and smoothness at $r= R_{\Lambda_\eta^\epsilon}^{(I)}$.
The strength $\kappa_8$ of the color-Coulomb potential is the same
for all $\bm{8}$-potentials. The $\bm{8}$-meson energy $E_{J^{PC}}^{(I)}$
is the same for all potentials associated with a given $\bm{8}$-meson.
We choose the relaxation length in the large-$r$ potential to be the Sommer scale:  $d = r_0 = 0.5$\,fm.
Parametrizations of 8 of the lowest B\nobreakdash-O potentials in pure $\mathrm{SU(3)}$ gauge theory 
were presented in Ref.~\cite{Ala24}.
We take their values for $\kappa_8 = 0.037$ and for $r_0^3 A_{\Lambda_\eta^\epsilon}$: 
0.11, 1.18, and 0.68 for $\Sigma_g^{+\prime}$, $\Pi_g$, and $\Sigma_u^-$.
We treat the $\bm{8}$-meson energies $E_{J^{PC}}^{(I)}$ as adjustable parameters.

\textbf{Schr\"odinger equation.}
The Schr\"odinger equation for a heavy-quark pair in the coupled $\Sigma_g^{+\prime}$ and $\Pi_g$ potentials 
and in the $\Sigma_u^-$ potential are given in Ref.~\cite{Ber24}.
We take the charm and bottom quark masses to be $m_c = 1.48$\,GeV  and $m_b = 4.89$\,GeV.
In the isospin-0 channel, we ignore the narrow avoided crossing with the $\Sigma_g^+$ quarkonium potential.

The exotic $c \bar{c}$ meson $\chi_{c1}(3872)$ ($X_c$) discovered in 2003 \cite{Cho03}
 is remarkable because its mass is within 100\,keV of the $D^{\ast0} \bar{D}^0$ threshold  \cite{PDG24}.
It has isospin 0 and $J^{PC} = 1^{++}$.
We identify $X_c$ with the $1^{++}$ state in the ground-state multiplet for the isospin-0 $1^{--}$ $\bm{8}$-meson.
The ground-state energy $\varepsilon_1^{(0)}$ depends on $E_{1^{--}}^{(0)}$.
There is a critical  energy $E_{1^{--}}^{(0)\ast}$ for which it is exactly at threshold: $\varepsilon_1^{(0)\ast} = 0$.
The critical energy from solving the Schr\"odinger equation with $m_c = 1.48$\,GeV is
$E_{1^{--}}^{(0)\ast}= -157$\,MeV.  

If a potential has the critical depth for the $c \bar{c}$ system,
 it must support $b \bar{b}$ bound states. 
Given $E_{1^{--}}^{(0)\ast}$, we determine their energies by solving the Schr\"odinger equation with $m_b = 4.89$\,GeV. 
The ground-state  energy is
$\varepsilon_1^{(0)} = -17$\,MeV. 
There is one other bound state with  energy $-4$~MeV.
Its multiplet is a doublet consisting of a spin-singlet $0^{-+}$ and a spin-triplet $1^{--}$.

The exotic $b \bar{b}$ mesons $Z_b$ and $Z_b^\prime$ are remarkable 
because their masses are only about 3\,MeV above the $B^\ast \bar{B}$ and $B^\ast \bar{B}^\ast$ thresholds. 
They have isospin 1 and the neutral mesons  are $1^{+-}$.
Two candidates for these two states are the $1^{+-}$ states in the ground-state multiplets
for the isospin-1 $1^{--}$ and $0^{-+}$ $\bm{8}$-mesons. 
The ground-state energies $\varepsilon_1^{(1)}$ and $\varepsilon_0^{(1)}$ 
depend on  $E_{1^{--}}^{(1)}$ and $E_{0^{-+}}^{(1)}$. 
There are critical energies $E_{1^{--}}^{(1)\ast}$ and $E_{0^{-+}}^{(1)\ast}$ 
for which both  ground states are exactly at threshold: $\varepsilon_1^{(1)\ast} = \varepsilon_0^{(1)\ast} = 0$.
The critical energies from solving the Schr\"odinger equation with $m_b = 4.89$\,GeV are
$E_{1^{--}}^{(1)\ast} = - 95$\,MeV and  $E_{0^{-+}}^{(1)\ast} = - 107$\,MeV.

These critical energies suggest that the lowest energy $\bm{8}$-hadron is an $\bm{8}$-meson with isospin 0 and $J^{PC}=1^{--}$. The $\bm{8}$-mesons with isospin 1 and $J^{PC}=0^{-+}$ and $1^{--}$ are higher in energy and close to each other. These predictions present a challenge for lattice QCD.

\textbf{Spin splittings.}
Quantitative predictions from  the  B\nobreakdash-O approximation require taking into account 
the spin and flavor splittings of heavy hadrons.
The spin splitting $\Delta$  between the $1^-$ and $0^-$ heavy mesons $M^\ast$ and $M$ scales 
with the heavy quark mass $m_Q$ as $1/m_Q$.
The spin splitting is $\Delta_c= 141$\,MeV for charm mesons and $\Delta_b= 45$\,MeV  for bottom mesons.
The energies of $M$ and $M^\ast$ relative to their spin average are $-\frac{3}{4} \Delta$ and  $+\frac{1}{4} \Delta$.
The thresholds for heavy-meson pairs are $-\frac{3}{2} \Delta$ for $M \bar{M}$,
$-\frac{1}{2} \Delta$ for $M^\ast\bar{M}$  and $M \bar{M}^\ast$, and $+\frac{1}{2} \Delta$ for $M^\ast\bar{M}^\ast$.

The B\nobreakdash-O potential matrix includes terms of order $1/m_Q$
that break the heavy-quark spin symmetry. 
They remove the degeneracies between states with different $J^{PC}$  in a B\nobreakdash-O multiplet
and mix states with the same $J^{PC}$ from different B-O multiplets.
At large $r$,  the  spin-splitting terms reduce to a constant potential matrix $V_\mathrm{SS}$
that depends on the spin vectors $\bm{S}_1$ and $\bm{S}_2$ of the heavy quark and antiquark:
\begin{equation}
V_\text{SS}(\bm{S}_1,\bm{S}_2)=\Delta \, \bm{j}_1 \cdot \bm{S}_1 + \Delta \, \bm{j}_2 \cdot \bm{S}_2,
\label{vss}
\end{equation}
where $\bm{j}_1$ and $\bm{j}_2$ are the light QCD angular momenta of the well-separated $\bm{3}$-meson and $\bm{3}^\ast$-meson. To determine its qualitative effects, we will take into account $V_\mathrm{SS}$ at first order in perturbation theory. This approximation puts the heavy-meson pair thresholds at their correct physical values.

We first consider the ground-state multiplet for a $1^{--}$ $\bm{8}$-meson.
If its energy in the absence of spin splittings is $\varepsilon_1$, the energies of the four states to order $\Delta$ are 
\begin{subequations}
\begin{alignat}{2}
&E_{1^{+-}} = \varepsilon_1, && E_{0^{++}} = \varepsilon_1-\Delta,
\label{E1--:1+-,0++}
\\
&E_{1^{++}} = \varepsilon_1-\tfrac{1}{2} \Delta, \qquad && E_{2^{++}} = \varepsilon_1+\tfrac{1}{2} \Delta.
\label{E1--:1+=,2++}
\end{alignat}
\end{subequations}
If  $\varepsilon_1$ is tuned to 0,
the $1^{++}$ and $2^{++}$ states remain near the $M^\ast \bar{M}$  and $M^\ast \bar{M}^\ast$ thresholds. 
The $0^{++}$ state is halfway between the $M \bar{M}$ and $M^\ast \bar{M}$ thresholds.
The $1^{+-}$ state is halfway between the $M^\ast \bar{M}$ and $M^\ast \bar{M}^\ast$ thresholds.

We next consider the ground-state multiplet for a  $0^{-+}$ $\bm{8}$-meson.
If its energy in the absence of spin splittings is $\varepsilon_0$,
the energies of the two states to order  $\Delta$ are 
\begin{equation}
E_{1^{+-}} =  E_{0^{++}} = \varepsilon_0.
\label{E0-+:1+-,0++}
\end{equation}
There are no corrections at first order in $\Delta$.
If $\varepsilon_0$ is tuned to 0,
the two states remain degenerate with energies halfway between the $M^\ast \bar{M}$ and $M^\ast \bar{M}^\ast$ thresholds.

If $\varepsilon_0$ and $\varepsilon_1$ are nearly degenerate, it is necessary to diagonalize the $2 \times 2$
submatrices of  $V_\mathrm{SS}$ for the $0^{++}$ and $1^{+-}$ states in the two multiplets.
The resulting pairs of energies  can be approximated by
\begin{subequations}
\begin{align}
E_{0^{++}}^\pm &= \frac{\varepsilon_1 + \varepsilon_0  - \Delta
 \pm \sqrt{( \varepsilon_1 - \varepsilon_0 - \Delta)^2 + 3 \Delta^2}}{2},
\label{E-0++pm}
\\
E_{1^{+-}}^\pm &= \frac{\varepsilon_1 + \varepsilon_0  
 \pm \sqrt{( \varepsilon_1 - \varepsilon_0)^2 + \Delta^2}}{2}.
\label{E-1+-pm}
\end{align}
\end{subequations}
If $\Delta^2$ is set to 0, these energies reduce to those in Eqs.~\eqref{E1--:1+-,0++} and \eqref{E0-+:1+-,0++}.
In the limit $\varepsilon_0 \to \varepsilon_1$,
the energies of the $0^{++}$ states are $\varepsilon_1 - \frac{3}{2} \Delta$ and $\varepsilon_1 + \frac{1}{2} \Delta$
and  those of the $1^{+-}$ states are $\varepsilon_1 - \frac{1}{2} \Delta$ and $\varepsilon_1 + \frac{1}{2} \Delta$.
If $\varepsilon_1$ and  $\varepsilon_0$ are both tuned to 0,
the two $0^{++}$ states are at the $M \bar{M}$ and $M^\ast \bar{M}^\ast$ thresholds
and the two $1^{+-}$ states are at the $M^\ast \bar{M}$ and $M^\ast \bar{M}^\ast$ thresholds.

The two $0^{++}$ states with energies in  Eq.~\eqref{E-0++pm}
and the two $1^{+-}$ states with  energies in  Eq.~\eqref{E-1+-pm} are both superpositions of 
heavy-quark-pair spin-singlet and spin-triplet states
in the ground-state multiplets of the $0^{-+}$   and $1^{--}$ $\bm{8}$-mesons.
For the $1^{+-}$ states,
the mixing angle $\theta_1$ satisfies $\tan(2\theta_1) = \Delta/(\varepsilon_1 - \varepsilon_0)$. 
For the $0^{++}$ states, the mixing angle $\theta_0$  satisfies 
$\tan(2\theta_0) = \sqrt{3} \Delta/(\varepsilon_1 - \varepsilon_0 - \Delta)$. 
In the limit $\varepsilon_0 \to \varepsilon_1$,  the two $1^{+-}$ states 
are equal-amplitude superpositions of spin-singlet  and spin-triplet.
The two $0^{-+}$ states are superpositions of spin-singlet  and spin-triplet with amplitudes that differ by a factor of $\sqrt{3}$.

Note that the limit $\varepsilon_0 \to \varepsilon_1$ reproduces a molecular model for the tetraquark mesons \cite{Vol16}. In general, the B\nobreakdash-O framework reproduces a molecular model in cases where the energy levels closest to the threshold are degenerate for all B\nobreakdash-O potentials that approach the same ($\bm{3}$+$\bm{3}^\ast$)-potential.

\begin{figure}
\centering
\includegraphics{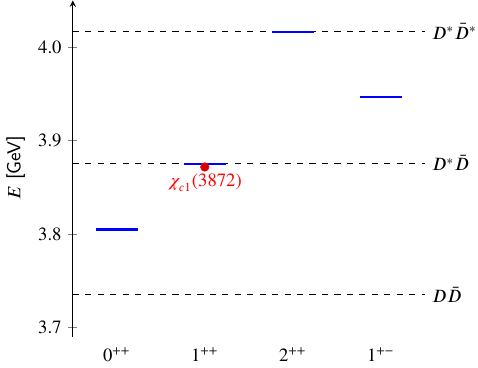}
\caption{%
Energy levels for the multiplet of isospin-0 hidden-charm tetraquark mesons associated with the $1^{--}$
$\bm{8}$-meson. The mass of $\chi_{c1}(3872)$ was used as an input to determine the energy of the
$\bm{8}$-meson. The dashed horizontal lines are the thresholds for charm-meson pairs.}
\label{fig:levelsX}
\end{figure}

\textbf{$\bm{X_c}$ and $\bm{X_b}$ Multiplets.}
If $E_{1^{--}}^{(0)}$ is tuned to its critical value in the $c \bar{c}$ system,  
we can predict the masses of $X_c$ and its B\nobreakdash-O partners to first order in $\Delta_c$.
The mass of $X_c$ is near the $D^\ast\bar{D}$ threshold at 3.876\,GeV. 
The mass of the $2^{++}$ partner is near the $D^\ast\bar{D}^\ast$ threshold at 4.017\,GeV.
The masses of the $0^{++}$ and $1^{+-}$ partners are near 3.805\,GeV and 3.946\,GeV.
The energy levels in the multiplet including $X_c$ are illustrated in Fig.~\ref{fig:levelsX}.

If $E_{1^{--}}^{(0)}$ is tuned to its critical value in the $c \bar{c}$ system,  
we can predict the masses for the ground-state multiplet of the $b \bar{b}$ system to first order in $\Delta_b$.
The mass of the $1^{++}$ state $X_b$ is about 17\,MeV below the $B^\ast \bar{B}$ threshold. 
The mass of the $2^{++}$ partner is about 17\,MeV  below the $B^\ast \bar{B}^\ast$ threshold. 
The masses of the $0^{++}$ and $1^{+-}$ partners are near 10.565\,GeV and 10.610\,GeV.

\begin{figure}
\centering
\includegraphics{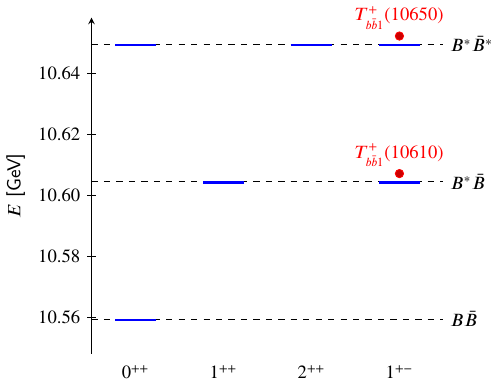}
\caption{%
Energy levels for the multiplets of isospin-1 hidden-bottom tetraquark mesons associated with the
$1^{--}$ and $0^{-+}$ $\bm{8}$-mesons. The masses of $T_{b\bar{b}1}^+(10610)$ and $T_{b\bar{b}1}^+(10650)$  were
used as inputs to determine the energies of the $\bm{8}$-mesons.
The dashed horizontal lines are the thresholds for bottom-meson pairs.}
\label{fig:levelsZ}
\end{figure}

\textbf{$\bm{Z_b}$ and $\bm{Z_b^\prime}$ Multiplets.}
Another remarkable feature of $Z_b^\prime$ besides its proximity to the $B^\ast \bar{B}^\ast$ threshold
is that, although it can decay into $B^\ast \bar{B}$ and $B \bar{B}^\ast$ through $S$-wave channels, these decays are not seen.
The suppression of these decays can be explained by a fine-tuning of $E_{1^{--}}^{(1)}$ and $E_{0^{-+}}^{(1)}$ 
so that $\varepsilon_1^{(1)} = \varepsilon_0^{(1)}$  in the $b \bar{b}$ system.
This equality would follow from a \emph{light-quark spin symmetry}  proposed by Voloshin \cite{Vol16}.
It implies that $Z_b^\prime$ is an equal-amplitude superposition of spin-singlet  and spin-triplet,
which prevents its decay into $B^\ast \bar{B}$ or $B \bar{B}^\ast$.
Masses of $Z_b$ and  $Z_b^\prime$  near the $B^\ast \bar{B}$  and  $B^\ast \bar{B}^\ast$ thresholds
can be explained by a further fine-tuning of
$E_{1^{--}}^{(1)}$ and $E_{0^{-+}}^{(1)}$ so that $\varepsilon_1^{(1)} = \varepsilon_0^{(1)} = 0$.
Given these fine tunings, we can predict the masses of $Z_b$ and  $Z_b^\prime$ 
and their B\nobreakdash-O partners  to first order in $\Delta_b$.
The $Z_b$ and  $Z_b^\prime$ are near the $B^\ast \bar{B}$ and $B^\ast \bar{B}^\ast$ thresholds.
The $1^{++}$ and $2^{++}$  partners are also near the $B^\ast \bar{B}$ and $B^\ast \bar{B}^\ast$ thresholds. 
The two $0^{++}$  partners are near the $B \bar{B}$ and $B^\ast \bar{B}$ thresholds. The energy levels in the multiplets including $Z_b$ and $Z_b^\prime$ are illustrated in Fig.~\ref{fig:levelsZ}.

\textbf{Discussion.} 
A B\nobreakdash-O potential for hidden-heavy hadrons that approaches a $\bm{8}$-potential
at small $r$
must connect to a ($\bm{3}$+$\bm{3}^\ast$)-potential at large $r$.
Only those $\bm{8}$-potentials associated with the lowest-energy $\bm{8}$-hadrons
cross below
the ($\bm{3}$+$\bm{3}^\ast$)-threshold before approaching it, in which case 
they can support bound states and/or resonances, which we identify as exotic hidden-heavy hadrons.
This explains why most exotic hidden-heavy hadrons have energies near heavy-hadron-pair thresholds.

This also provides insight into the remarkable properties of some of the hidden-heavy tetraquark mesons.
The proximity of $X_c$ to the $D^\ast\bar{D}$ threshold comes from
the fine tuning of an isospin-0 $\bm{8}$-meson energy.
The suppression of decays of  $Z_b^\prime$ into $B^\ast \bar{B}$  and  $B \bar{B}^\ast$ and
the proximity of  $Z_b$  and  $Z_b^\prime$ to the $B^\ast \bar{B}$  and  $B^\ast \bar{B}^\ast$ thresholds
come from fine tunings of  two  isospin-1 $\bm{8}$-meson energies. 

We estimated the energies of the B\nobreakdash-O  partners of $X_c$, $X_b$, $Z_b$, and $Z_b^\prime$ 
to first order in $\Delta$.
Accurate predictions would require solving the coupled-channel Schr\"odinger equations
including the spin-splitting matrix $V_\mathrm{SS}$.
In the isospin-0 case, it is also necessary to take into account the narrow avoided crossing 
between the $\Sigma_g^+$ and $\Sigma_g^{+\prime}$ potentials \cite{Bra24}.

Similar methods can be applied to hidden-heavy pentaquark baryons.
They are bound states or resonances in $\bm{8}$-potentials 
that cross below the threshold for the pair of a $\bm{3}$-baryon and a $\bm{3}^\ast$-meson
before approaching it.
The ground-state pentaquark multiplet could be a
triplet associated with a $J^P=\frac{1}{2}^+$ $\bm{8}$-baryon or a quartet
associated with a $\frac{3}{2}^+$ $\bm{8}$-baryon \cite{Ber24}. 

The continuity constraint has important implications for lattice-QCD calculations of B\nobreakdash-O potentials.
The operators used previously in lattice QCD with light quarks produce states at small $r$ 
with very small  overlaps with $\bm{8}$-hadrons 
\cite{Bal05,Bal00,Alb17,Bul19,Bic20,Pre20,Sad21,Bul24}.  
Much larger overlaps would be obtained with operators that connect the $\bm{3} $ and $\bm{3}^\ast$ sources 
by Wilson lines to a junction with a color-octet QCD operator. 
Until lattice-QCD calculations of the B\nobreakdash-O potentials are available,
B\nobreakdash-O calculations of exotic hidden-heavy hadrons will have to rely on model  potentials.
Since $\bm{8}$-potentials must connect to  ($\bm{3}$+$\bm{3}^\ast$)-potentials, 
any model reduces  to the interpolation between the large-$r$ and small-$r$ regions.
This dramatically reduces the model dependence of  the predictions.

Physics beyond the Standard Model may include long-lived heavy particles with $\bm{8}$ color charges,
such as the gluino in some supersymmetric models.
After being created in a collider, a gluino can hadronize before it decays.
The ATLAS and CMS collaborations have searched for long-lived gluinos at the Large Hadron Collider  
\cite{CMS11,ATLAS11a,ATLAS11b,CMS12,ATLAS13,ATLAS16,ATLAS19,CMS19}. 
The analyses depend on assumptions about the gluino-hadron spectrum 
\cite{Bae99,Hew04,Arv07,Far11,Rab15},
which is the $\bm{8}$-hadron spectrum up to an additive constant associated with the gluino mass.
The exotic hidden-heavy hadrons provide constraints on that spectrum.
The masses of $X_c$ and $Z_b$ suggest that the lightest gluino-meson is isospin-0 but isospin-1 is not much heavier.
The masses of the $c \bar{c}$ pentaquarks $P_{c\bar{c}}^+(4312)$ discovered in 2019 \cite{LHCB19}
and $P_{c\bar{c} s}^+(4338)$ discovered in 2023 \cite{LHCB23}
suggest that the lightest gluino-baryon is isospin-$\frac{1}{2}$ but  isospin-0  is not much heavier.
Quantitative B\nobreakdash-O analyses of exotic hidden-heavy hadrons
that take into account heavy-hadron spin splittings could determine the lowest-energy states 
in the $\bm{8}$-hadron spectrum and  enable  more accurate constraints on long-lived gluinos.

\begin{acknowledgments}
This work was supported in part by the U.S. Department of Energy under grant DE-SC0011726.
EB would like to thank A.~Mohapatra for illuminating discussions that led to the realization
that $\bm{8}$-potentials must connect to ($\bm{3}$+$\bm{3}^\ast$)-potentials.
EB would also like to thank N.~Brambilla and A.~Vairo for hospitality and discussions
during two extended visits to Technical University of Munich.
This work contributes to the goals of the US DOE ExoHad Topical Collaboration, Contract DE-SC0023598.
\end{acknowledgments}

\bibliography{xhhhbib}
	
\end{document}